\documentclass[english,aps,superscriptaddress,preprintnumbers,reprint,footinbib,amsmath,amssymb,prb,longbibliography]{revtex4-2}
\usepackage[latin9]{inputenc}
\usepackage{color}
\usepackage{amsmath}
\usepackage{amssymb}
\usepackage{graphicx}
\usepackage{esint}

\makeatletter

\@ifundefined{textcolor}{}{%
 \definecolor{BLACK}{gray}{0}
 \definecolor{WHITE}{gray}{1}
 \definecolor{RED}{rgb}{1,0,0}
 \definecolor{GREEN}{rgb}{0,1,0}
 \definecolor{BLUE}{rgb}{0,0,1}
 \definecolor{CYAN}{cmyk}{1,0,0,0}
 \definecolor{MAGENTA}{cmyk}{0,1,0,0}
 \definecolor{YELLOW}{cmyk}{0,0,1,0}
}

\usepackage{aecompl}

\usepackage{epsfig}\usepackage{dcolumn}\usepackage{bm}
\usepackage{babel}

\usepackage{hyperref}
\hypersetup{
    colorlinks=true,
    linkcolor=blue,
    urlcolor=blue,
    citecolor=blue
        }

\makeatother

\usepackage{babel}
\begin{document}
\title{Impact of Broken Inversion Symmetry on Molecular States in multi-Weyl
fermions}
\author{W.C. Silva}
\email[corresponding author: ]{willian.carvalho@unesp.br}

\affiliation{São Paulo State University (Unesp), School of Engineering, Department
of Physics and Chemistry, 15385-007, Ilha Solteira-SP, Brazil}
\author{J.E. Sanches}
\affiliation{São Paulo State University (Unesp), School of Engineering, Department
of Physics and Chemistry, 15385-007, Ilha Solteira-SP, Brazil}
\author{D.S. Rojo}
\affiliation{São Paulo State University (Unesp), School of Engineering, Department
of Physics and Chemistry, 15385-007, Ilha Solteira-SP, Brazil}
\author{L. Squillante}
\affiliation{São Paulo State University (Unesp), IGCE, Department of Physics, 13506-970,
Rio Claro-SP, Brazil}
\author{\\
 M. de Souza}
\affiliation{São Paulo State University (Unesp), IGCE, Department of Physics, 13506-970,
Rio Claro-SP, Brazil}
\author{M.S. Figueira}
\affiliation{Instituto de Física, Universidade Federal Fluminense, 24210-340, Niterói,
Rio de Janeiro, Brazil}
\author{I.A. Shelykh}
\affiliation{Science Institute, University of Iceland, Dunhagi-3, IS-107, Reykjavik,
Iceland}
\author{E. Marinho Jr.}
\email[corresponding author: ]{enesio.marinho@unesp.br}

\affiliation{São Paulo State University (Unesp), School of Engineering, Department
of Physics and Chemistry, 15385-007, Ilha Solteira-SP, Brazil}
\author{A.C. Seridonio}
\email[corresponding author: ]{antonio.seridonio@unesp.br}

\affiliation{São Paulo State University (Unesp), School of Engineering, Department
of Physics and Chemistry, 15385-007, Ilha Solteira-SP, Brazil}
\begin{abstract}
We study inversion-symmetry (IS) breaking in impurity dimers coupled
to topological multi-Weyl systems in the low-energy dispersion domain.
In the IS-preserved multi-Weyl semimetal phase, Hubbard bands split
into \textit{symmetric} and \textit{antisymmetric} \textit{molecular-like}
subbands. Breaking IS induces a transition to a multi-Weyl metal,
lifting the degeneracy of the Weyl node and closing the \textit{pseudogap}.
This causes opposite energy shifts: valence-band \textit{symmetric}
(\textit{antisymmetric}) subbands red- (blue-) shift, reversing in
the conduction band until a degeneracy point. Beyond this threshold,
\textit{symmetric} bands flatten near band cutoffs, whereas \textit{antisymmetric}
bands form quasi-zero energy modes asymptotically approaching- yet
never crossing- the Fermi level. Crucially, identical molecular symmetries
maintain nondegeneracy even as energy separation vanishes with stronger
IS breaking. Our results demonstrate symmetry-selective mechanisms
for topological molecular states in multi-Weyl systems.
\end{abstract}
\maketitle

\section{Introduction}

Multi-Weyl semimetals~\cite{Fang,Dantas,Huang,Liu} represent a topological
generalization of conventional Weyl semimetals~\cite{Armitage,Hu,Hasan2021,Hasan2017,Yan,Zheng},
being distinguished by higher topological charges ($J>1$) and associated
phenomena such as chiral anomalies, nonlinear optical responses, topological
Fano interference, anomalous thermoelectric transport, and Hall effects~\cite{Hayata,Huang2,Bharti,Ahn,Mukherjee,Park,TopoFano,Thermo1,Thermo2,Chen,Xu,Xiong,Dantas2018,Transport,NonlinearQT}.
These systems reveal highly anisotropic low-energy dispersions with
relativistic linear scaling ($E\propto\pm v_{F}k_{z}$) along a high-symmetry
axis, and quadratic or higher-power law scaling ($E\propto\pm\alpha k_{\perp}^{J}$)
in the orthogonal plane ($k_{\perp}=\sqrt{k_{x}^{2}+k_{y}^{2}}$),
and $J$ denotes the integer topological charge~\cite{Dantas}, related
to the Chern number $C=\pm J$ through the Berry flux $\Phi_{B}=\oint_{S}\mathbf{\Omega}\cdot d\mathbf{S}=s2\pi J$,
where $\mathbf{\Omega}$ is the Berry curvature and $s=\pm1$ are
the chirality indexes~\cite{Hasan2021}. Multi-Weyl nodes manifest
themselves as momentum-space monopole-antimonopole pairs with quantized
Berry fluxes $\pm2\pi J$. Their opposite chiralities act as momentum-space
sources and sinks, connected via topologically protected Fermi arc
surface states~\cite{Hasan2021,Hasan2017}.

The topological charge $J$ originates from the merging of $J$ co-chiral
Weyl nodes under $C_{2J}$ rotational symmetry protection ($J\leq3$)~\cite{Fang}.
Furthermore, the bulk-boundary correspondence enforces $J$ Fermi
arcs per surface projection~\cite{Dantas}. The prototypical realizations
include $\text{HgCr}_{2}\text{Se}_{4}$ and $\text{SrSi}_{2}$ (double-Weyl,
$J=2$)~\cite{Fang,Xu,Huang,Chen} and $\text{A(MoX)}_{3}$ ($\text{A = Rb, Tl}$;
$\text{X = Te}$, triple-Weyl, $J=3$)~\cite{Liu}. These materials
exemplify the interplay between crystalline symmetries ($C_{4}$,
$C_{6}$) and band topology, stabilizing exotic quasiparticles (multi-Weyl
fermions) beyond the Standard Model and providing platforms to explore
quantum anomalies and correlated topological phases.

\begin{figure}[ht]
\centering\includegraphics[width=1\columnwidth]{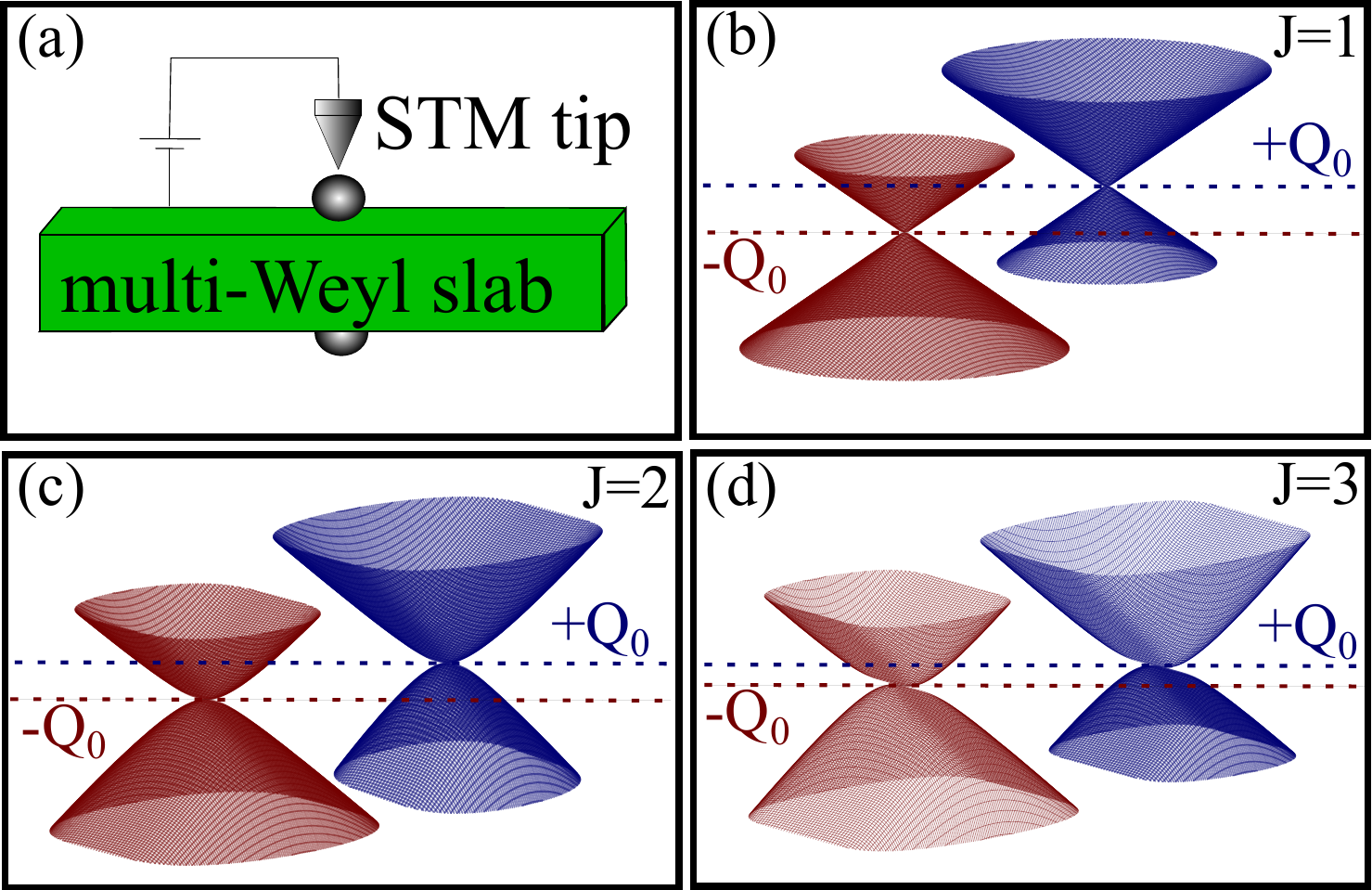}\caption{\label{fig:Pic1} {(Color online) (a) {A multi-Weyl
slab \cite{NonlinearQT}} featuring a dimer of side-coupled impurities
as the prototype system for observing our findings: a specific IS-breaking
perturbation $Q_{0}\protect\neq0$ induces degeneracy between the
centers of \textit{symmetric} and\textit{ antisymmetric} Hubbard \textit{molecular-like}
subbands. Beyond the threshold, the \textit{antisymmetric} centers
migrate toward the Fermi level without crossing it, forming a pair
of quasi-zero energy modes, while the \textit{symmetric} counterparts
shift diametrically away, as in Figs.\ref{fig:Pic2} and \ref{fig:Pic3}.
{The STM tip is expected to access the }\textit{{molecular-like}}{{}
subbands in the impurity density of states.} Band structures corresponding
to (b) $J=1$ (single), (c) $J=2$ (double) and (d) $J=3$ (triple)
Weyl metals exhibiting broken IS. }}
\end{figure}

A promising route to probe emergent phenomena in multi-Weyl systems
is by the introduction of magnetic impurities, as this reveals certain
fundamental mechanisms governed by these quasiparticles, in particular
the synergy between topology and strong electron correlations. Key
examples include Kondo single-impurity and lattice systems~\cite{Pedrosa,Miranda,KondoLattice1,KondoLattice3},
as well as the two-impurity problem characterized by the Ruderman-Kittel-Kasuya-Yosida
(RKKY) mediated interaction~\cite{RKKY1,RKKY2,RKKY3,RKKY4,RKKY5}.
These systems demonstrate how the enhanced topological charges of
multi-Weyl semimetals reshape well-established many-body phenomena.

In this work, we extend previous studies that use impurities as topological
probes by exploring emergent regimes beyond the Kondo and RKKY frameworks.
We focus on the Coulomb blockade regime~\cite{Hubbard1963,TopoFano,Flensberg,HubbardI},
investigating a dimer of impurities coupled to inversion symmetry
(IS)-broken topological multi-Weyl systems in the domain of low energy
excitations {[}Fig.\ref{fig:Pic1}{]}, where novel quantum molecular
states appear.

Starting with the case of intact IS, we observe the formation of \textit{symmetric}
and \textit{antisymmetric} Hubbard \textit{molecular-like} subbands.
Breaking of the IS shifts Weyl nodes, closes the \textit{pseudogap}
characteristic for the semimetallic regime, and drives the system
into a Weyl metal phase with renormalized subbands. The \textit{symmetric}
(\textit{antisymmetric}) subband undergoes redshift (blueshift) in
the valence band and blueshift (redshift) in the conduction band until
degeneracy at a specific IS-breaking threshold. Beyond this threshold,
\textit{symmetric} subbands shift away from the Fermi level, revealing
a pronounced broadening, while \textit{antisymmetric} subbands, playing
the role of quasi-zero energy modes, asymptotically approaching the
Fermi level without crossing it. Importantly, degeneracy arises exclusively
between subbands of opposite symmetry, underscoring the symmetry-selective
behavior inherent to multi-Weyl systems.

\section{The Model}

{To analyze a dimer mediated by low-energy multi-Weyl
fermions ($\mathcal{H}_{\text{{mW}}}$), we propose the slab setup
\cite{NonlinearQT} shown in Fig.\ref{fig:Pic1}(a). }We describe
the system by the Hamiltonian (setting $\hbar=1$),
\begin{equation}
\mathcal{H}=\mathcal{H}_{\text{mW}}+\mathcal{H}_{\text{dimer}},
\end{equation}
which is consistent with the Anderson-like framework \cite{Anderson,TopoFano}.

{The first term accounts for the slab geometry \cite{NonlinearQT}
\begin{equation}
\mathcal{H}_{\text{{mW}}}=\sum_{\textbf{k}s}\psi_{\textbf{k}s}^{\dagger}s[D(\tilde{k}_{-}^{J}\sigma_{+}+\tilde{k}_{+}^{J}\sigma_{-})+v_{F}k_{zs}\sigma_{z}-\sigma_{0}Q_{0}]\psi_{\textbf{k}s}\label{eq:mWeyl}
\end{equation}
and describes itinerant states via the spinor $\psi_{\textbf{k}s}^{\dagger}=(c_{\textbf{k}s\uparrow}^{\dagger}\;c_{\textbf{k}s\downarrow}^{\dagger})$,
where $c_{\textbf{k}s\sigma}^{\dagger}$ ($c_{\textbf{k}s\sigma}$)
creates (annihilates) an electron with momentum $\textbf{k}$, spin
$\sigma=\uparrow,\downarrow$ and chirality $s=\pm1$ for Weyl nodes
at energies $sQ_{0}$ and $k_{zs}=k_{z}-sQ$ breaks time reversal
symmetry (TRS).} Here, $\tilde{k}_{\pm}=(k_{x}\pm ik_{y})/k_{D}$
and $J$ denotes the winding number, with $k_{D}=D/v_{F}$ ($\pm D$)
representing the \textit{Debye-like} momentum (infrared/ultraviolet)
band cutoff. The Pauli matrices are given by $\sigma_{\pm}=\frac{1}{2}(\sigma_{x}\pm i\sigma_{y})$
and $\sigma_{z}$, $\sigma_{0}$ is the identity matrix, the velocity
$v_{F}$ defines the slope of the Weyl cone along $k_{z}$.

The topological charge $sJ$, related to the Berry curvature flux,
ensures the appearance of $J$ Fermi arc pairs at the material boundaries
due to the bulk-boundary correspondence \cite{Dantas}. The dispersion
relation $\varepsilon_{\textbf{k}s}^{\pm}=\pm\varepsilon_{\textbf{k}s}-sQ_{0}$
($\varepsilon_{\textbf{k}s}=v_{F}\sqrt{k_{zs}^{2}+|\tilde{k}_{+}|^{2J}k_{D}^{2}}$)
splits into the conduction ($+$) and valence ($-$) bands, as shown
in Figs.\ref{fig:Pic1}(b)-(d).

The dimer Hamiltonian reads
\begin{align}
\mathcal{H}_{\text{dimer}} & =\sum_{j\sigma}(\varepsilon_{d}+\frac{U}{2})n_{j\sigma}+\frac{U}{2}(\sum_{j\sigma}n_{j\sigma}-1)^{2}-\frac{U}{2}\nonumber \\
 & +\frac{1}{\sqrt{\mathcal{N}}}\sum_{j\textbf{k}\sigma s}(v_{\textbf{k}j}d_{j\sigma}^{\dagger}c_{\textbf{k}s\sigma}+\text{H.c.}),\label{eq:Imp}
\end{align}
{where $\varepsilon_{d}$ ($j=1,2$) corresponds to
the impurity energies,} $n_{j\sigma}=d_{j\sigma}^{\dagger}d_{j\sigma}$
is the number operator, $U$ is the on-site Coulomb repulsion, and
$v_{\textbf{k}j}$ quantifies atom-host coupling.

The system's spectral properties derive from impurity ($\text{DOS}_{jj}$)
and molecular (\textit{symmetric/antisymmetric}) orbital densities
of states. Using the equation-of-motion method \cite{Flensberg} for
Green's functions (GFs), we analyze the particle-hole symmetric (PHS)
regime by: i) setting {$\varepsilon_{d}=-\frac{U}{2}$
}to nullify the first term of Eq.(\ref{eq:Imp});{{}
ii) assuming $v_{\textbf{k}j}\approx v$ (local coupling).}

\begin{figure*}[ht]
\centering\includegraphics[width=0.99\textwidth,height=0.4\textheight]{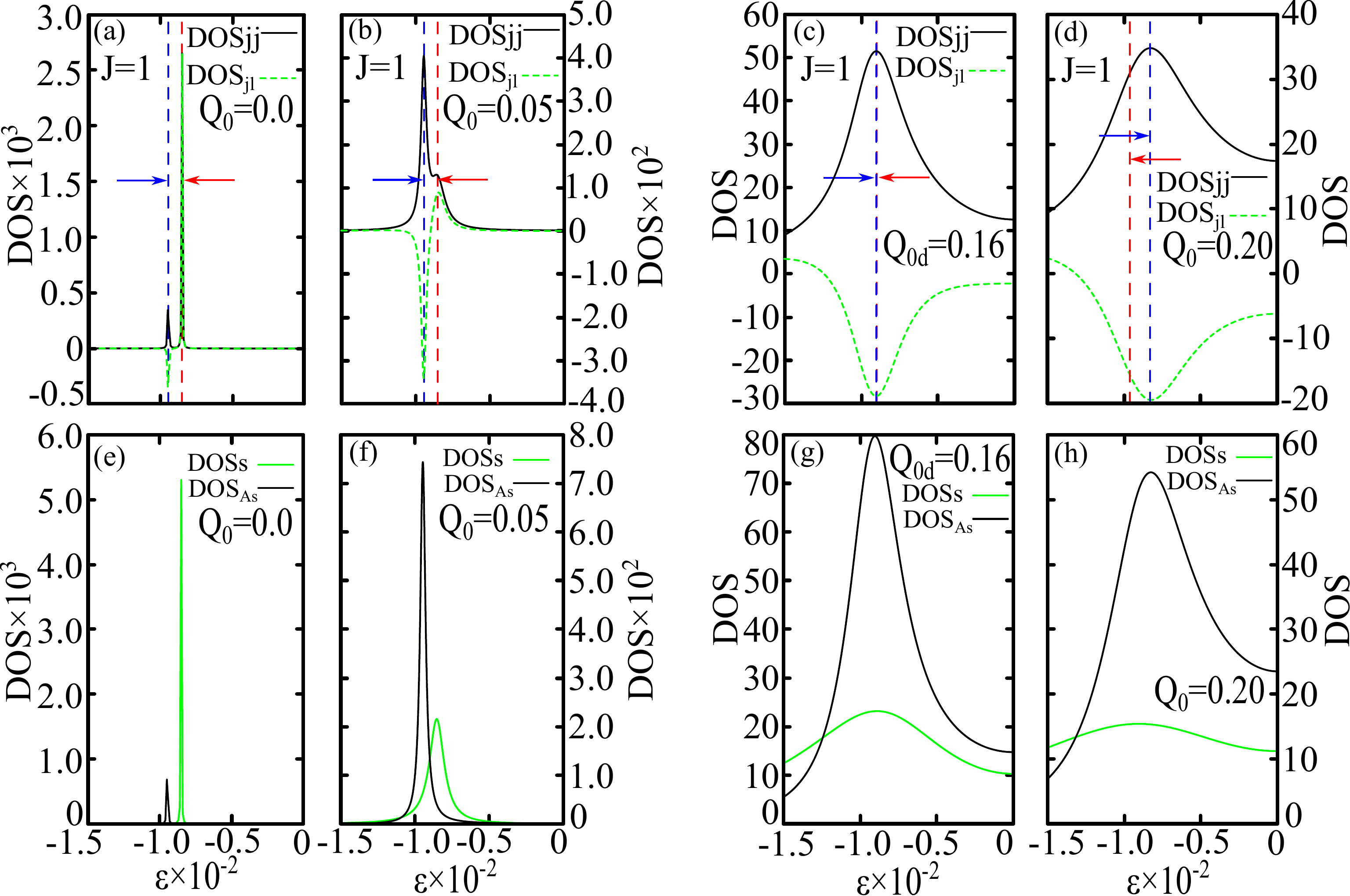}\caption{\label{fig:Pic2} {(Color online) Single Weyl PHS analysis: (a)-(d)
Diagonal $\text{DOS}_{jj}$ and off-diagonal $\text{DOS}_{jl}$ spectra
in the valence band (negative energies), comparing preserved IS in
(a) with progressively broken IS in (b)-(d). In (a), the \textit{antisymmetric}
(blue dashed line) and \textit{symmetric} (red dashed line) subbands
are clearly resolved. Weak IS breaking in (b) induces partial merging
of the subbands via redshift (\textit{symmetric}) and blueshift (\textit{antisymmetric}),
as indicated by the arrows. Full coalescence into a degenerate \textit{molecular-like}
state occurs at the specific threshold in (c). Beyond this threshold
(d), spectral broadening flattens the subband features. (e)-(h) Corresponding
\textit{antisymmetric} $\text{DOS}_{\text{As}}$ and \textit{symmetric}
$\text{DOS}_{\text{S}}$ profiles under increasing IS breaking. Notably,
panel (h) demonstrates a loss of resonance between the subbands as
they shift out of alignment.}}
\end{figure*}

These assumptions render $\mathcal{H}$ invariant under particle-hole
transformations $c_{\textbf{k}s\sigma}\rightarrow c_{-\textbf{k}s\sigma}^{\dagger}$
and $d_{j\sigma}\rightarrow-d_{j\sigma}^{\dagger}$, yielding $\langle n_{j\sigma}\rangle=\frac{1}{2}$
and giving mirror-symmetric spectral profiles. The impurity DOS, $\text{DOS}_{jl}=(-1/\pi)\text{Im}\tilde{\mathcal{G}}_{jl}$
($j,l=1,2$), is given by the time Fourier transform (TFT) of the
GF $\mathcal{G}_{jl}=-i\theta(\tau)\langle\{d_{j\sigma}(\tau),d_{l\sigma}^{\dagger}(0)\}\rangle_{\mathcal{H}}$.

In the Coulomb blockade regime, when the Kondo and RKKY correlations
are irrelevant \cite{KondoWeyl1,Pedrosa,Miranda,KondoLattice1,KondoLattice3,RKKY1,RKKY2,RKKY3,RKKY4},
one can safely use the Hubbard-I approximation \cite{Hubbard1963,TopoFano,Flensberg,HubbardI}.
{This implies that a more sophisticated decoupling
scheme suffices to capture the strong correlation effects arising
from impurity spins, the detailed implementation of which we defer
to future work. Such a solution is expected to elucidate novel spin
dynamics mechanisms in molecular structures embedded in multi-Weyl
fermion systems.} In this way, the diagonal GF ($j=l$) becomes
\begin{align}
\tilde{\mathcal{G}}_{jj} & =\frac{\frac{1}{2}}{[\tilde{g}_{jj}(\varepsilon+\frac{U}{2})]^{-1}-\tilde{g}_{jj}(\varepsilon+\frac{U}{2})[\tilde{t}(\varepsilon-\frac{U}{2})]^{2}}\nonumber \\
 & +\frac{\frac{1}{2}}{[\tilde{g}_{jj}(\varepsilon-\frac{U}{2})]^{-1}-\tilde{g}_{jj}(\varepsilon-\frac{U}{2})[\tilde{t}(\varepsilon+\frac{U}{2})]^{2}},\label{eq:Gmolecule}
\end{align}
where $\tilde{g}_{jj}(\varepsilon\pm\frac{U}{2})=[\varepsilon\pm\frac{U}{2}-v^{2}\tilde{\mathcal{G}}_{\sigma}^{0}]^{-1}$
corresponds to single GFs at Hubbard peaks $\varepsilon=\pm\frac{U}{2}$,
with
\begin{equation}
\tilde{\mathcal{G}}_{\sigma}^{0}=\sum_{s}\tilde{\mathcal{G}}_{\sigma s}^{0}=\frac{1}{\mathcal{N}}\sum_{\textbf{k}s}\frac{\varepsilon_{s}+i0^{+}}{[\varepsilon_{s}+i0^{+}]^{2}-\varepsilon_{\textbf{k}s}^{2}}\label{eq:GFtilfo2}
\end{equation}
being the multi-Weyl fermion GF ($\varepsilon_{s}=\varepsilon+sQ_{0}$).

Notably, the Hubbard peaks at $\varepsilon=-\frac{U}{2}$ (first term)
and $\varepsilon=\frac{U}{2}$ (second term) in Eq.(\ref{eq:Gmolecule})
split into \textit{symmetric} and \textit{antisymmetric} \textit{molecular-like}
states (the subbands centers) due to the effective hopping
\begin{equation}
\tilde{t}(\varepsilon\pm\frac{U}{2})=[1\mp\frac{U}{2}\tilde{g}_{jj}(\varepsilon\pm\frac{U}{2})]v^{2}\tilde{\mathcal{G}}_{\sigma}^{0},\label{eq:complexh}
\end{equation}
which depends on $\tilde{\mathcal{G}}_{\sigma}^{0}$ and $U$. The
off-diagonal GF ($j\neq l$)
\begin{equation}
\tilde{\mathcal{G}}_{jl}=\tilde{t}(\varepsilon\pm\frac{U}{2})\tilde{g}_{jj}(\varepsilon+\frac{U}{2})\tilde{\mathcal{G}}_{ll}\label{eq:CrossedGF}
\end{equation}
encodes quantum interference between dimer impurities.

For comparison, consider a conventional dimer with atomic energy $\varepsilon_{0}$
and hopping $t$: $\mathcal{H}_{\text{dimer}}=\varepsilon_{0}\sum_{j\sigma}n_{j\sigma}+\sum_{\sigma}(td_{1\sigma}^{\dagger}d_{2\sigma}+\text{H.c.})$.
Here, Eq.(\ref{eq:Gmolecule}) reduces to $\tilde{\mathcal{G}}_{jj}=[\varepsilon-\varepsilon_{0}+i0^{+}-(\varepsilon-\varepsilon_{0}+i0^{+})^{-1}|t|^{2}]^{-1}$,
yielding molecular energies $\varepsilon-\varepsilon_{0}=\pm|t|$
(where $-$/$+$ corresponds to \textit{symmetric}/\textit{antisymmetric}
states). Analogously, Eq.(\ref{eq:Gmolecule}) for our system gives
\begin{equation}
\varepsilon+\frac{U}{2}-\text{Re}[v^{2}\tilde{\mathcal{G}}_{\sigma}^{0}]=\pm\text{Re}[\tilde{t}(\varepsilon-\frac{U}{2})]\label{eq:Ed}
\end{equation}
and
\begin{equation}
\varepsilon-\frac{U}{2}-\text{Re}[v^{2}\tilde{\mathcal{G}}_{\sigma}^{0}]=\pm\text{Re}[\tilde{t}(\varepsilon+\frac{U}{2})]\label{eq:EdPlusU}
\end{equation}
for split Hubbard subbands centers near $\varepsilon=-\frac{U}{2}$
and $\varepsilon=\frac{U}{2}$, respectively.

The molecular densities of states $\text{DOS}_{\text{S(As)}}=(-1/\pi)\text{Im}\tilde{\mathcal{G}}_{\text{S(As)}}$
can be derived from the TFT of $\mathcal{G}_{\text{S(As)}}=-i\theta(\tau)\langle\{\frac{1}{\sqrt{2}}(d_{1\sigma}(\tau)\pm d_{2\sigma}(\tau)),\frac{1}{\sqrt{2}}(d_{1\sigma}^{\dagger}(0)\pm d_{2\sigma}^{\dagger}(0))\}\rangle_{\mathcal{H}}$,
leading to
\begin{equation}
\text{DOS}_{\text{S(As)}}=\frac{1}{2}\left[\text{DOS}_{11}+\text{DOS}_{22}\pm(\text{DOS}_{12}+\text{DOS}_{21})\right],\label{eq:DOSbab}
\end{equation}
where $+$ ($-$) denotes \textit{symmetric} (\textit{antisymmetric})
linear combination of the atomic orbitals of the impurities.

To compute $\tilde{\mathcal{G}}_{\sigma}^{0}$ (Eq.\ref{eq:GFtilfo2}),
we use the \textit{Debye-like} sphere approximation $\mathcal{N}=\sum_{\textbf{k}s}=\frac{\Omega}{(2\pi)^{3}}\int d^{3}\textbf{k}=\frac{\Omega}{6\pi^{2}}k_{D}^{3}$
and hyperspherical coordinates: $k_{x}=k_{D}(\frac{\varepsilon_{\textbf{k}s}\sin\theta}{D})^{1/J}\cos\phi$,
$k_{y}=k_{D}(\frac{\varepsilon_{\textbf{k}s}\sin\theta}{D})^{1/J}\sin\phi$,
$k_{zs}=k_{D}\frac{\varepsilon_{\textbf{k}s}}{D}\cos\theta$ with Jacobian
$\text{J}(\varepsilon_{\textbf{k}s},\theta,\phi)=\frac{k_{D}^{3}}{D}(\frac{\varepsilon_{\textbf{k}s}}{D})^{2/J}\frac{(\sin\theta)^{\frac{2}{J}-1}}{J}$
and $\int\tilde{\mathcal{G}}_{\sigma}^{0}d^{3}\textbf{k}=\int\tilde{\mathcal{G}}_{\sigma}^{0}\text{{J}}(\varepsilon_{\textbf{k}s},\theta,\phi)d\varepsilon_{\textbf{k}s}d\theta d\phi$
\cite{TopoFano}. {This yields
\begin{align}
\text{Im}\tilde{\mathcal{G}}_{\sigma}^{0} & =\sum_{s}\text{Im}\tilde{\mathcal{G}}_{\sigma s}^{0}=-\pi\rho_{0}\nonumber \\
 & =-\sum_{s}\frac{3\pi^{3/2}\Gamma(\frac{1}{J})}{4JD^{\frac{J+2}{J}}\Gamma(\frac{2+J}{2J})}\varepsilon_{s}^{2/J}\varTheta(|D|-\varepsilon_{s}),\label{eq:pristineLDOS}
\end{align}
where $\varGamma(x)$ and $\varTheta(x)$ are the Gamma and Heaviside
step functions, respectively. }{The power law scaling of $\varepsilon_{s}^{2/J}$
reflects the pristine DOS $\rho_{0}$, which is metallic (finite at
the Fermi level) when IS is broken but exhibits a characteristic semimetallic
\textit{pseudogap} otherwise.}{{} Due to the restriction
$\varTheta(|D|-\varepsilon_{s})$, we ensure that the model parameters
remain within the validity range where the density of states $\rho_{0}$
accurately describes the low-energy multi-Weyl fermionic states. These
states are confined within the }\textit{{Debye}}{{}
sphere of radius $k_{D}$.}

{Using the Kramers-Kronig relation, we obtain
\begin{align}
\text{Re}\tilde{\mathcal{G}}_{\sigma s}^{0} & =-\frac{1}{\pi}\int_{-D}^{+D}\frac{\text{Im}\tilde{\mathcal{G}}_{\sigma s}^{0}}{\varepsilon_{s}-y}dy\nonumber \\
 & =-\frac{1}{\pi}\text{{sgn}}(\varepsilon_{s})\text{{Im}}\tilde{\mathcal{G}}_{\sigma s}^{0}\text{{P.V.}}\int_{-D/\varepsilon_{s}}^{+D/\varepsilon_{s}}K(u)du,\label{eq:KramersK}
\end{align}
wherein $y=\varepsilon_{s}u,$ \text{P.V.} stands for the Cauchy
principal value and $K(u)=\frac{(u^{2})^{1/J}}{1-u}$ is a kernel
function. To properly evaluate the integral over $u$ in Eq.(\ref{eq:KramersK}),
we first observe that $K(u)$ remains non-vanishing in the asymptotic
limits $u\rightarrow\pm\infty$ (corresponding to the wide-band limit
$\pm D/\varepsilon_{s}\rightarrow\pm\infty$) for $J=1,2$. This challenge
is resolved through analytical integration while maintaining a fixed
ratio $D/\varepsilon_{s}$, yielding}

{{}
\begin{equation}
\text{Re}\tilde{\mathcal{G}}_{\sigma}^{0}(J=1)=\sum_{s}\frac{3\varepsilon_{s}}{D^{3}}(\frac{1}{2}\varepsilon_{s}\ln\frac{|D+\varepsilon_{s}|}{|D-\varepsilon_{s}|}-D)\label{eq:J1x}
\end{equation}
and
\begin{equation}
\text{Re}\tilde{\mathcal{G}}_{\sigma}^{0}(J=2)=\sum_{s}\frac{3\pi}{8D^{2}}\varepsilon_{s}\ln\frac{\varepsilon_{s}^{2}}{|\varepsilon_{s}^{2}-D^{2}|}.\label{eq:J2x}
\end{equation}
}

{In the case $J=3,$ the $K(u)$ vanishing behavior
at $u\rightarrow\pm\infty$ allows the safe replacement $\int_{-D/\varepsilon_{s}}^{+D/\varepsilon_{s}}K(u)du\rightarrow\int_{-\infty}^{+\infty}K(u)du$
in Eq.(\ref{eq:KramersK}), achieving }

{{}
\begin{equation}
\text{Re}\tilde{\mathcal{G}}_{\sigma}^{0}(J=3)=\sum_{s}\text{sgn}(\varepsilon_{s})\tan(C_{2J})\text{Im}\tilde{\mathcal{G}}_{\sigma s}^{0},\label{eq:J3x}
\end{equation}
where $C_{2J}\equiv360^{\circ}/2J$ reflects the rotational symmetry
stabilizing multi-Weyl points \cite{TopoFano,Fang}. As a consequence,
in contrast to ordinary metals, the wide-band limit ($\varepsilon_{s}/D\ll1,$
equivalently) does not yield a purely imaginary and energy-independent
self-energy, $v^{2}\tilde{\mathcal{G}}_{\sigma}^{0}\approx-i\pi v^{2}\rho_{0}=-i\Gamma$,
when discrete states couple to Weyl quasiparticles.}

\section{Results and Discussion}

Within the PHS regime, where $\varepsilon_{d}=-\frac{U}{2}$ and $v_{\mathbf{k}j}\approx v$,
it is sufficient to analyze the spectral properties in the negative
energy domain (valence band), as the positive energy region (conduction
band) exhibits mirror-symmetric characteristics. Consequently, we
further discuss only the first partial fraction in Eq.(\ref{eq:Gmolecule})
of $\tilde{\mathcal{G}}_{jj}$ corresponding to the valence band.
We fix the parameters to $\varepsilon_{d}=-0.01D$, $U=0.02D$, $v=0.14D$,
and $D=1$ throughout the work.

We first examine the single Weyl case ($J=1$), which displays the
diagonal $\text{DOS}_{jj}=(-1/\pi)\text{Im}\,\tilde{\mathcal{G}}_{jj}$
and off-diagonal $\text{DOS}_{jl}=(-1/\pi)\text{Im}\,\tilde{\mathcal{G}}_{jl}$
as IS is progressively broken, see Figs.\ref{fig:Pic2}(a)-(d). The
double ($J=2$) and triple ($J=3$) Weyl cases exhibit qualitatively
similar behavior, making their explicit presentation unnecessary.
The vertical dashed lines in such panels mark the centers of the \textit{molecular-like}
subbands, whose positions are determined by numerically solving Eq.(\ref{eq:Ed})
for the poles of the GF $\tilde{\mathcal{G}}_{jj}$ {[}Eq.(\ref{eq:Gmolecule}){]}.

\begin{figure}[ht]
\centering\includegraphics[width=1\columnwidth]{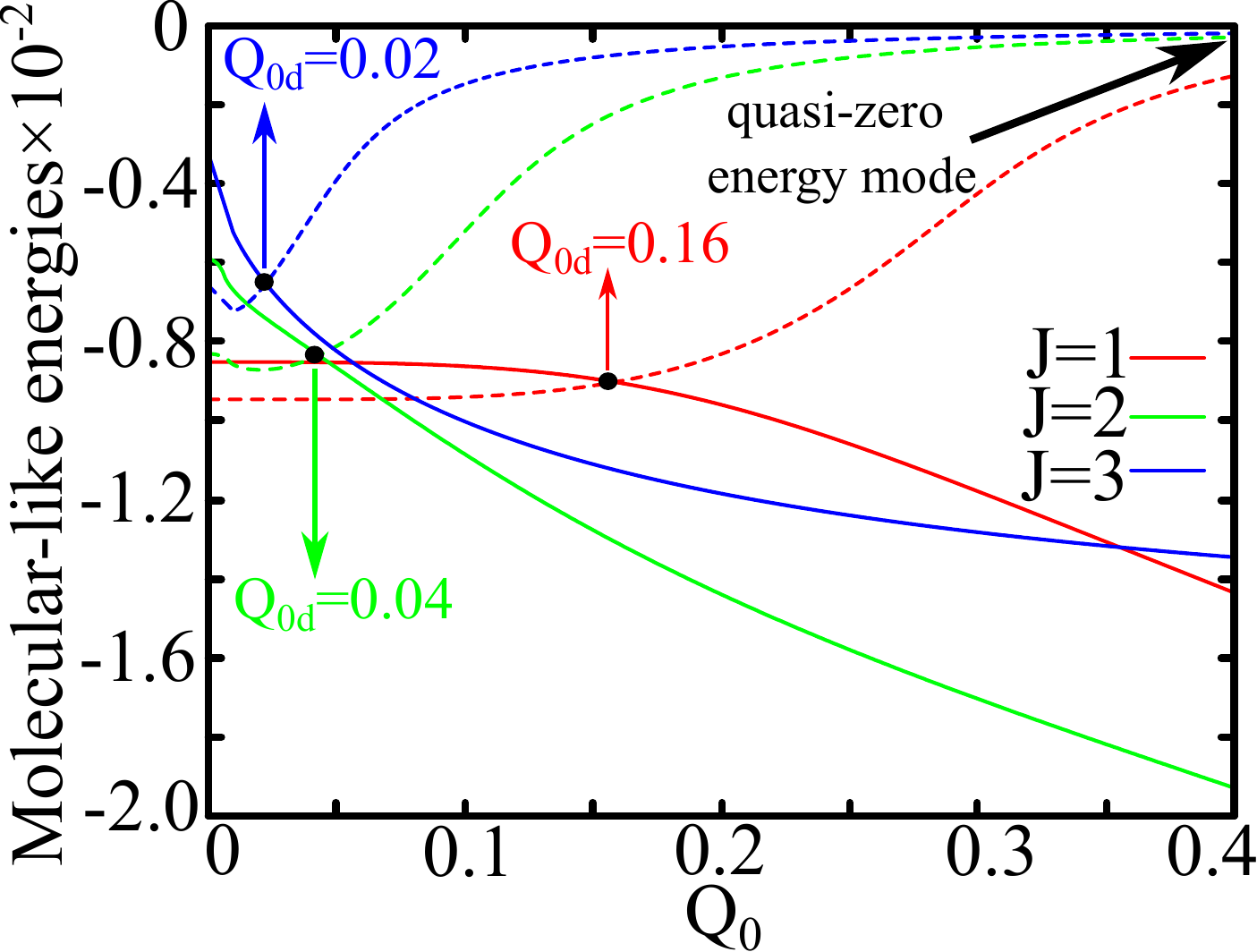} \caption{\label{fig:Pic3} {(Color online) Numerical \textit{molecular-like}
energies from Eq.(\ref{eq:Ed}) as a function of IS breaking parameter
$Q_{0}$ for single ($J=1$), double ($J=2$) and triple ($J=3$)
Weyl fermions in the negative energy domain (valence band). Mirror-symmetric
behavior governs the conduction band. The specific thresholds $Q_{0\text{d}}$
mark the IS breaking strengths where the \textit{symmetric} and \textit{antisymmetric}
subband centers achieve degeneracy. For $Q_{0}>Q_{0\text{d}}$, the
\textit{symmetric} subband center shifts away from the Fermi level,
while the \textit{antisymmetric} center asymptotically approaches
it. Orbitals with identical symmetry types (e.g., \textit{symmetric/symmetric}
or \textit{antisymmetric/antisymmetric}) are prohibited from degenerating,
resulting in two quasi-zero energy modes flanking the Fermi level.}}
\end{figure}

\begin{figure*}[ht]
\centering\includegraphics[width=0.98\textwidth,height=0.3\textheight]{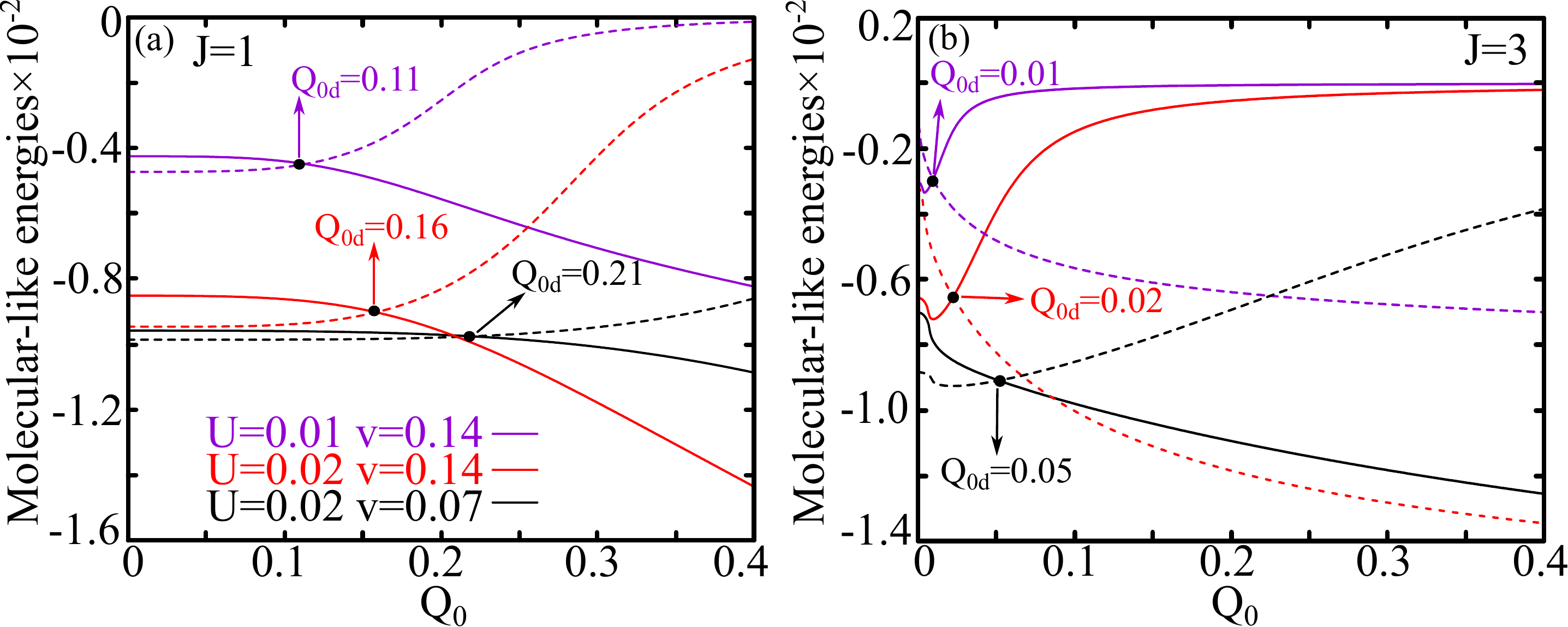}\caption{\label{fig:Pic4} {{(Color online) }\textit{{Molecular-like}}{{}
energies with $Q_{0}$ for $J=1$ (a) and $J=3$ (b) multi-Weyl hosts
in the valence band for several values of Coulomb repulsion $U$ and
atom-host coupling $v$. The $J=2$ case (not shown) follows analogous
trend.} }}
\end{figure*}

In panel (a) with intact IS ($Q_{0}=0$), $\text{DOS}_{jl}$ reveals
split Hubbard \textit{molecular-like} subbands in the multi-Weyl semimetal
regime: an \textit{antisymmetric} dip (blue dashed line) and a \textit{symmetric}
peak (red dashed line) in the vicinity of the original band center
$\varepsilon_{d}=-0.01D$. For $\text{DOS}_{jj}$, the \textit{antisymmetric}
subband center corresponds to a peak. This splitting arises from the
term $\tilde{t}(\varepsilon-\frac{U}{2})$ in Eq.(\ref{eq:complexh}),
which introduces energy-dependent hopping mediated by multi-Weyl quasiparticles
via $\tilde{\mathcal{G}}_{\sigma}^{0}$ {[}Eq.(\ref{eq:GFtilfo2}){]}.
In contrast, conventional molecules are characterized by static hopping
(determined by wavefunction overlap), which precludes state degeneracy.

When IS is weakly broken ($Q_{0}=0.05D$, panel (b)), the \textit{pseudogap}
in Eq.(\ref{eq:pristineLDOS}) for $\rho_{0}$ closes due to the energy
shift of the Weyl nodes, signaling the transition to the multi-Weyl
metallic phase. Consequently, $\text{DOS}_{jj}$ and $\text{DOS}_{jl}$
display blue- and redshifted \textit{antisymmetric} and \textit{symmetric}
subbands, respectively (see the arrows in Fig.\ref{fig:Pic2}). These
subbands broaden and merge, fully coalescing at $Q_{0\text{d}}=0.16D$
(panel (c)), where their centers align resonantly at the degeneracy
point. Here, $\text{DOS}_{jj}$ exhibits a single peak, while $\text{DOS}_{jl}$
retains a dip. {In experimental settings, an application
of a stress serves as a means of achieving these novel states as it
breaks IS \cite{Armitage}. Furthermore, differential conductance
measurements using an STM tip {[}Fig.\ref{fig:Pic1}(a){]} right above
one of the impurities (atomic limit) can provide direct access to
the $\text{DOS}_{jj},$ where the }\textit{{symmetric}}{{}
and }\textit{{antisymmetric}}{{}
subbands are present {[}Figs.\ref{fig:Pic2}(a)-(d){]}. }For $Q_{0}>Q_{0\text{d}}$
(e.g., $Q_{0}=0.20D$, panel (d)), increased broadening obscures the
separation between subband centers (vertical dashed lines).

Panels (e)-(h) present $\text{DOS}_{\text{S}}$ and $\text{DOS}_{\text{As}}$
from Eq.(\ref{eq:DOSbab}), which characterize the \textit{symmetric}
and \textit{antisymmetric} spectral densities, respectively, in the
presence of Weyl quasiparticles. These panels follow trends analogous
to those of (a)-(d). {We expect that these molecular
limits could be experimentally accessible by depositing impurities
in close proximity on the same surface of the multi-Weyl slab, with
the STM tip positioned directly above one impurity and laterally to
the other. In this configuration, the tip-impurity couplings may interfere
constructively (destructively), enabling selective probing of either
$\text{DOS}_{\text{S}}$ or $\text{DOS}_{\text{As}}$ {[}Figs.\ref{fig:Pic2}(e)-(h){]}.
This effect stems from quantum interference between such tunneling
channels.}

At the threshold $Q_{0\text{d}}=0.16D$, $\text{DOS}_{\text{As}}$
at the degeneracy point becomes significantly stronger than $\text{DOS}_{\text{S}}$.
The enhancement of the amplitude of this resonant state originates
from the off-diagonal $\text{DOS}_{jl}$ at this energy, which acquires
a large negative value {[}panel (c){]}. For $\text{DOS}_{\text{As}}$,
the interference term $-(\text{DOS}_{12}+\text{DOS}_{21})$ amplifies
spectral weight, whereas for $\text{DOS}_{\text{S}}$, the sign-flipped
term $+(\text{DOS}_{12}+\text{DOS}_{21})$ suppresses it.

Above the threshold ($Q_{0}>Q_{0\text{d}}$), stronger IS break enhances
spectral broadening in the \textit{ antisymmetric} (blueshifted) and
\textit{symmetric} (redshifted) components {[}panel (h){]}, with the
latter being more pronounced. Simultaneously, the \textit{symmetric}
center shifts toward the lower (infrared $-D$) band cutoff, while
the \textit{antisymmetric} component approaches the Fermi energy from
below. In the conduction band (not shown), this trend reverses: \textit{antisymmetric}
(redshifted) and \textit{symmetric} (blueshifted) components shift
toward the Fermi energy and upper (ultraviolet $+D$) band cutoff,
respectively.

To elucidate this behavior, we explicitly show in Fig.\ref{fig:Pic3}
the numerical solutions of Eq.(\ref{eq:Ed}) for the $Q_{0}$- and
$J$-dependent \textit{antisymmetric} and \textit{symmetric} subband
centers for the cases of single ($J=1$), double ($J=2$) and triple
($J=3$) Weyl fermions. This approach systematically tracks the evolution
of the subband center as the IS is broken. Particularly, the degeneracy
threshold $Q_{0}$ decreases with increasing $J$. For $J=3$, the
\textit{antisymmetric} level asymptotically approaches the Fermi energy
from below more rapidly than for $J=1,2$. Remarkably, \textit{antisymmetric}
centers persist near the Fermi energy as quasi-zero energy modes,
indicating that orbitals of identical symmetry (e.g., \textit{antisymmetric}-type)
cannot achieve degeneracy even when $Q_{0}$ is extremely close to
the energy cutoff $D.$ Conversely, \textit{symmetric}-type orbitals
move progressively toward the infrared and ultraviolet band cutoffs,
consistent with their suppressed spectral weight, and likewise fail
to degenerate.

{Notably, the threshold amplitude $Q_{0\text{d}}$
depends on $J$, the on-site Coulomb repulsion $U$, and the atom-host
coupling $v$. To elucidate these dependencies, we analyze the effective
hopping {[}Eq.(\ref{eq:Ed}), right-hand side{]}, which vanishes at
$Q_{0}=Q_{0\text{d}}$, i.e., when $\text{Re}[\tilde{t}(\varepsilon-\frac{U}{2})]=0$.
This condition coincides with the degeneracy of the }\textit{{antisymmetric}}{{}
and }\textit{{symmetric}}{{} subband
centers. Consequently, the degenerate }\textit{{molecular-like}}{{}
energy $\varepsilon$ satisfies $\varepsilon+\frac{U}{2}=\text{Re}[v^{2}\tilde{\mathcal{G}}_{\sigma}^{0}]$.
In the wide-band limit $\varepsilon_{s}/D\ll1$, the $J=1$ case of
Eq.(\ref{eq:J1x}) yields $\text{Re}[v^{2}\tilde{\mathcal{G}}_{\sigma}^{0}]\approx\frac{3v^{2}}{D}\sum_{s}[-\left(\frac{\varepsilon_{s}}{D}\right)+\left(\frac{\varepsilon_{s}}{D}\right)^{3}]\propto\frac{v^{2}}{D^{3}}\left(\frac{\varepsilon}{D}\right)Q_{0\text{d}}^{2},$
which is a quadratic polynomial in the inversion-symmetry-breaking
parameter $Q_{0\text{d}}$. This reveals that $Q_{0\text{d}}$ scales
directly with $U$ but inversely with $v$ {[}Fig.\ref{fig:Pic4}(a),
Eq.(\ref{eq:Ed}){]}. Similar trends hold for $J=2$ (not shown) and
$J=3$ {[}Fig.\ref{fig:Pic4}(b){]}. In summary, since $\text{Re}[v^{2}\tilde{\mathcal{G}}_{\sigma}^{0}]$
is governed by Eqs.(\ref{eq:J1x})-(\ref{eq:J3x}) in determining
the degenerate }\textit{{molecular-like}}{{}
energy $\varepsilon$ for fixed $U$ and $v$, the threshold $Q_{0\text{d}}$
exhibits a distinct $J$-dependence. This behavior is reminiscent
from the power-law structure of Eq.(\ref{eq:pristineLDOS}) for $\rho_{0}$.}

\section{Conclusions}

We demonstrated that \textit{symmetric} and \textit{antisymmetric}
\textit{molecular-like} subbands in a multi-Weyl semimetal dimer host
novel quantum states in the IS breaking regime. In the IS-preserved
multi-Weyl phase, these subbands remain distinct. At a specific IS-breaking
strength, the lifting of Weyl node degeneracy closes the \textit{pseudogap},
and drives the system to a multi-Weyl metallic phase where the subband
centers resonate. Beyond this threshold, \textit{symmetric} subbands
move toward infrared and ultraviolet band cutoffs, exhibiting spectral
broadening and flattening, while \textit{antisymmetric} subband centers
asymptotically approach the Fermi level without crossing it, forming
a pair of quasi-zero energy modes. These results provide deeper insights
into quasiparticle mediated chemical bonding mechanisms in multi-Weyl
systems, thereby highlighting a pathway to engineer novel molecular
states in topological Weyl materials.

\section{Acknowledgments}

We thank the Brazilian funding agencies CNPq (Grants. Nr. 302887/2020-2,
303772/2023-9, 311980/2021-0, and 308695/2021-6), the São Paulo Research
Foundation (FAPESP; Grant No. 2023/13467-6), Coordenação de Aperfeiçoamento
de Pessoal de Nível Superior - Brasil (CAPES) -- Finance Code 001
and FAPERJ process Nr. 210355/2018.


\begin{thebibliography}{39}%
\makeatletter
\providecommand \@ifxundefined [1]{%
 \@ifx{#1\undefined}
}%
\providecommand \@ifnum [1]{%
 \ifnum #1\expandafter \@firstoftwo
 \else \expandafter \@secondoftwo
 \fi
}%
\providecommand \@ifx [1]{%
 \ifx #1\expandafter \@firstoftwo
 \else \expandafter \@secondoftwo
 \fi
}%
\providecommand \natexlab [1]{#1}%
\providecommand \enquote  [1]{``#1''}%
\providecommand \bibnamefont  [1]{#1}%
\providecommand \bibfnamefont [1]{#1}%
\providecommand \citenamefont [1]{#1}%
\providecommand \href@noop [0]{\@secondoftwo}%
\providecommand \href [0]{\begingroup \@sanitize@url \@href}%
\providecommand \@href[1]{\@@startlink{#1}\@@href}%
\providecommand \@@href[1]{\endgroup#1\@@endlink}%
\providecommand \@sanitize@url [0]{\catcode `\\12\catcode `\$12\catcode
  `\&12\catcode `\#12\catcode `\^12\catcode `\_12\catcode `\%12\relax}%
\providecommand \@@startlink[1]{}%
\providecommand \@@endlink[0]{}%
\providecommand \url  [0]{\begingroup\@sanitize@url \@url }%
\providecommand \@url [1]{\endgroup\@href {#1}{\urlprefix }}%
\providecommand \urlprefix  [0]{URL }%
\providecommand \Eprint [0]{\href }%
\providecommand \doibase [0]{https://doi.org/}%
\providecommand \selectlanguage [0]{\@gobble}%
\providecommand \bibinfo  [0]{\@secondoftwo}%
\providecommand \bibfield  [0]{\@secondoftwo}%
\providecommand \translation [1]{[#1]}%
\providecommand \BibitemOpen [0]{}%
\providecommand \bibitemStop [0]{}%
\providecommand \bibitemNoStop [0]{.\EOS\space}%
\providecommand \EOS [0]{\spacefactor3000\relax}%
\providecommand \BibitemShut  [1]{\csname bibitem#1\endcsname}%
\let\auto@bib@innerbib\@empty
\bibitem [{\citenamefont {Fang}\ \emph {et~al.}(2012)\citenamefont {Fang},
  \citenamefont {Gilbert}, \citenamefont {Dai},\ and\ \citenamefont
  {Bernevig}}]{Fang}%
  \BibitemOpen
  \bibfield  {author} {\bibinfo {author} {\bibfnamefont {C.}~\bibnamefont
  {Fang}}, \bibinfo {author} {\bibfnamefont {M.~J.}\ \bibnamefont {Gilbert}},
  \bibinfo {author} {\bibfnamefont {X.}~\bibnamefont {Dai}},\ and\ \bibinfo
  {author} {\bibfnamefont {B.~A.}\ \bibnamefont {Bernevig}},\ }\bibfield
  {title} {\bibinfo {title} {Multi-Weyl topological semimetals stabilized by
  point group symmetry},\ }\href
  {https://doi.org/10.1103/physrevlett.108.266802} {\bibfield  {journal}
  {\bibinfo  {journal} {Phys. Rev. Lett.}\ }\textbf {\bibinfo {volume} {108}},\
  \bibinfo {pages} {266802} (\bibinfo {year} {2012})}\BibitemShut {NoStop}%
\bibitem [{\citenamefont {Dantas}\ \emph {et~al.}(2020)\citenamefont {Dantas},
  \citenamefont {Pe\~na Benitez}, \citenamefont {Roy},\ and\ \citenamefont
  {Sur\'owka}}]{Dantas}%
  \BibitemOpen
  \bibfield  {author} {\bibinfo {author} {\bibfnamefont {R.~M.~A.}\
  \bibnamefont {Dantas}}, \bibinfo {author} {\bibfnamefont {F.}~\bibnamefont
  {Pe\~na Benitez}}, \bibinfo {author} {\bibfnamefont {B.}~\bibnamefont
  {Roy}},\ and\ \bibinfo {author} {\bibfnamefont {P.}~\bibnamefont
  {Sur\'owka}},\ }\bibfield  {title} {\bibinfo {title} {Non-abelian anomalies
  in multi-Weyl semimetals},\ }\href
  {https://doi.org/10.1103/physrevresearch.2.013007} {\bibfield  {journal}
  {\bibinfo  {journal} {Phys. Rev. Research}\ }\textbf {\bibinfo {volume}
  {2}},\ \bibinfo {pages} {013007} (\bibinfo {year} {2020})}\BibitemShut
  {NoStop}%
\bibitem [{\citenamefont {Huang}\ \emph {et~al.}(2016)\citenamefont {Huang},
  \citenamefont {Xu}, \citenamefont {Belopolski}, \citenamefont {Lee},
  \citenamefont {Chang}, \citenamefont {Chang}, \citenamefont {Wang},
  \citenamefont {Alidoust}, \citenamefont {Bian}, \citenamefont {Neupane},
  \citenamefont {Sanchez}, \citenamefont {Zheng}, \citenamefont {Jeng},
  \citenamefont {Bansil}, \citenamefont {Neupert}, \citenamefont {Lin},\ and\
  \citenamefont {Hasan}}]{Huang}%
  \BibitemOpen
  \bibfield  {author} {\bibinfo {author} {\bibfnamefont {S.-M.}\ \bibnamefont
  {Huang}}, \bibinfo {author} {\bibfnamefont {S.-Y.}\ \bibnamefont {Xu}},
  \bibinfo {author} {\bibfnamefont {I.}~\bibnamefont {Belopolski}}, \bibinfo
  {author} {\bibfnamefont {C.-C.}\ \bibnamefont {Lee}}, \bibinfo {author}
  {\bibfnamefont {G.}~\bibnamefont {Chang}}, \bibinfo {author} {\bibfnamefont
  {T.-R.}\ \bibnamefont {Chang}}, \bibinfo {author} {\bibfnamefont
  {B.}~\bibnamefont {Wang}}, \bibinfo {author} {\bibfnamefont {N.}~\bibnamefont
  {Alidoust}}, \bibinfo {author} {\bibfnamefont {G.}~\bibnamefont {Bian}},
  \bibinfo {author} {\bibfnamefont {M.}~\bibnamefont {Neupane}}, \bibinfo
  {author} {\bibfnamefont {D.}~\bibnamefont {Sanchez}}, \bibinfo {author}
  {\bibfnamefont {H.}~\bibnamefont {Zheng}}, \bibinfo {author} {\bibfnamefont
  {H.-T.}\ \bibnamefont {Jeng}}, \bibinfo {author} {\bibfnamefont
  {A.}~\bibnamefont {Bansil}}, \bibinfo {author} {\bibfnamefont
  {T.}~\bibnamefont {Neupert}}, \bibinfo {author} {\bibfnamefont
  {H.}~\bibnamefont {Lin}},\ and\ \bibinfo {author} {\bibfnamefont {M.~Z.}\
  \bibnamefont {Hasan}},\ }\bibfield  {title} {\bibinfo {title} {New type of
  Weyl semimetal with quadratic double Weyl fermions},\ }\href
  {https://doi.org/10.1073/pnas.1514581113} {\bibfield  {journal} {\bibinfo
  {journal} {Proceedings of the National Academy of Sciences}\ }\textbf
  {\bibinfo {volume} {113}},\ \bibinfo {pages} {1180} (\bibinfo {year}
  {2016})}\BibitemShut {NoStop}%
\bibitem [{\citenamefont {Liu}\ and\ \citenamefont {Zunger}(2017)}]{Liu}%
  \BibitemOpen
  \bibfield  {author} {\bibinfo {author} {\bibfnamefont {Q.}~\bibnamefont
  {Liu}}\ and\ \bibinfo {author} {\bibfnamefont {A.}~\bibnamefont {Zunger}},\
  }\bibfield  {title} {\bibinfo {title} {Predicted realization of cubic Dirac
  fermion in quasi-one-dimensional transition-metal monochalcogenides},\ }\href
  {https://doi.org/10.1103/PhysRevX.7.021019} {\bibfield  {journal} {\bibinfo
  {journal} {Phys. Rev. X}\ }\textbf {\bibinfo {volume} {7}},\ \bibinfo {pages}
  {021019} (\bibinfo {year} {2017})}\BibitemShut {NoStop}%
\bibitem [{\citenamefont {Armitage}\ \emph {et~al.}(2018)\citenamefont
  {Armitage}, \citenamefont {Mele},\ and\ \citenamefont
  {Vishwanath}}]{Armitage}%
  \BibitemOpen
  \bibfield  {author} {\bibinfo {author} {\bibfnamefont {N.~P.}\ \bibnamefont
  {Armitage}}, \bibinfo {author} {\bibfnamefont {E.~J.}\ \bibnamefont {Mele}},\
  and\ \bibinfo {author} {\bibfnamefont {A.}~\bibnamefont {Vishwanath}},\
  }\bibfield  {title} {\bibinfo {title} {Weyl and Dirac semimetals in
  three-dimensional solids},\ }\href
  {https://doi.org/10.1103/revmodphys.90.015001} {\bibfield  {journal}
  {\bibinfo  {journal} {Rev. Mod. Phys.}\ }\textbf {\bibinfo {volume} {90}},\
  \bibinfo {pages} {015001} (\bibinfo {year} {2018})}\BibitemShut {NoStop}%
\bibitem [{\citenamefont {Hu}\ \emph {et~al.}(2019)\citenamefont {Hu},
  \citenamefont {Xu}, \citenamefont {Ni},\ and\ \citenamefont {Mao}}]{Hu}%
  \BibitemOpen
  \bibfield  {author} {\bibinfo {author} {\bibfnamefont {J.}~\bibnamefont
  {Hu}}, \bibinfo {author} {\bibfnamefont {S.-Y.}\ \bibnamefont {Xu}}, \bibinfo
  {author} {\bibfnamefont {N.}~\bibnamefont {Ni}},\ and\ \bibinfo {author}
  {\bibfnamefont {Z.}~\bibnamefont {Mao}},\ }\bibfield  {title} {\bibinfo
  {title} {Transport of topological semimetals},\ }\href
  {https://doi.org/10.1146/annurev-matsci-070218-010023} {\bibfield  {journal}
  {\bibinfo  {journal} {Annual Review of Materials Research}\ }\textbf
  {\bibinfo {volume} {49}},\ \bibinfo {pages} {207} (\bibinfo {year}
  {2019})}\BibitemShut {NoStop}%
\bibitem [{\citenamefont {Hasan}\ \emph {et~al.}(2021)\citenamefont {Hasan},
  \citenamefont {Chang}, \citenamefont {Belopolski}, \citenamefont {Bian},
  \citenamefont {Xu},\ and\ \citenamefont {Yin}}]{Hasan2021}%
  \BibitemOpen
  \bibfield  {author} {\bibinfo {author} {\bibfnamefont {M.~Z.}\ \bibnamefont
  {Hasan}}, \bibinfo {author} {\bibfnamefont {G.}~\bibnamefont {Chang}},
  \bibinfo {author} {\bibfnamefont {I.}~\bibnamefont {Belopolski}}, \bibinfo
  {author} {\bibfnamefont {G.}~\bibnamefont {Bian}}, \bibinfo {author}
  {\bibfnamefont {S.-Y.}\ \bibnamefont {Xu}},\ and\ \bibinfo {author}
  {\bibfnamefont {J.-X.}\ \bibnamefont {Yin}},\ }\bibfield  {title} {\bibinfo
  {title} {Weyl, Dirac and high-fold chiral fermions in topological quantum
  matter},\ }\href {https://doi.org/10.1038/s41578-021-00301-3} {\bibfield
  {journal} {\bibinfo  {journal} {Nature Reviews Materials}\ }\textbf {\bibinfo
  {volume} {6}},\ \bibinfo {pages} {784} (\bibinfo {year} {2021})}\BibitemShut
  {NoStop}%
\bibitem [{\citenamefont {Hasan}\ \emph {et~al.}(2017)\citenamefont {Hasan},
  \citenamefont {Xu}, \citenamefont {Belopolski},\ and\ \citenamefont
  {Huang}}]{Hasan2017}%
  \BibitemOpen
  \bibfield  {author} {\bibinfo {author} {\bibfnamefont {M.~Z.}\ \bibnamefont
  {Hasan}}, \bibinfo {author} {\bibfnamefont {S.-Y.}\ \bibnamefont {Xu}},
  \bibinfo {author} {\bibfnamefont {I.}~\bibnamefont {Belopolski}},\ and\
  \bibinfo {author} {\bibfnamefont {S.-M.}\ \bibnamefont {Huang}},\ }\bibfield
  {title} {\bibinfo {title} {Discovery of Weyl fermion semimetals and
  topological fermi arc states},\ }\href
  {https://doi.org/10.1146/annurev-conmatphys-031016-025225} {\bibfield
  {journal} {\bibinfo  {journal} {Annual Review of Condensed Matter Physics}\
  }\textbf {\bibinfo {volume} {8}},\ \bibinfo {pages} {289} (\bibinfo {year}
  {2017})}\BibitemShut {NoStop}%
\bibitem [{\citenamefont {Yan}\ and\ \citenamefont {Felser}(2017)}]{Yan}%
  \BibitemOpen
  \bibfield  {author} {\bibinfo {author} {\bibfnamefont {B.}~\bibnamefont
  {Yan}}\ and\ \bibinfo {author} {\bibfnamefont {C.}~\bibnamefont {Felser}},\
  }\bibfield  {title} {\bibinfo {title} {Topological materials: Weyl
  semimetals},\ }\href
  {https://doi.org/10.1146/annurev-conmatphys-031016-025458} {\bibfield
  {journal} {\bibinfo  {journal} {Annual Review of Condensed Matter Physics}\
  }\textbf {\bibinfo {volume} {8}},\ \bibinfo {pages} {337} (\bibinfo {year}
  {2017})}\BibitemShut {NoStop}%
\bibitem [{\citenamefont {Zheng}\ and\ \citenamefont
  {Zahid~Hasan}(2018)}]{Zheng}%
  \BibitemOpen
  \bibfield  {author} {\bibinfo {author} {\bibfnamefont {H.}~\bibnamefont
  {Zheng}}\ and\ \bibinfo {author} {\bibfnamefont {M.}~\bibnamefont
  {Zahid~Hasan}},\ }\bibfield  {title} {\bibinfo {title} {Quasiparticle
  interference on type-i and type-ii Weyl semimetal surfaces: a review},\
  }\href {https://doi.org/10.1080/23746149.2018.1466661} {\bibfield  {journal}
  {\bibinfo  {journal} {Advances in Physics: X}\ }\textbf {\bibinfo {volume}
  {3}},\ \bibinfo {pages} {1466661} (\bibinfo {year} {2018})}\BibitemShut
  {NoStop}%
\bibitem [{\citenamefont {Hayata}\ \emph {et~al.}(2017)\citenamefont {Hayata},
  \citenamefont {Kikuchi},\ and\ \citenamefont {Tanizaki}}]{Hayata}%
  \BibitemOpen
  \bibfield  {author} {\bibinfo {author} {\bibfnamefont {T.}~\bibnamefont
  {Hayata}}, \bibinfo {author} {\bibfnamefont {Y.}~\bibnamefont {Kikuchi}},\
  and\ \bibinfo {author} {\bibfnamefont {Y.}~\bibnamefont {Tanizaki}},\
  }\bibfield  {title} {\bibinfo {title} {Topological properties of the chiral
  magnetic effect in multi-Weyl semimetals},\ }\href
  {https://doi.org/10.1103/physrevb.96.085112} {\bibfield  {journal} {\bibinfo
  {journal} {Phys. Rev. B}\ }\textbf {\bibinfo {volume} {96}},\ \bibinfo
  {pages} {085112} (\bibinfo {year} {2017})}\BibitemShut {NoStop}%
\bibitem [{\citenamefont {Huang}\ \emph {et~al.}(2017)\citenamefont {Huang},
  \citenamefont {Zhou},\ and\ \citenamefont {Shen}}]{Huang2}%
  \BibitemOpen
  \bibfield  {author} {\bibinfo {author} {\bibfnamefont {Z.-M.}\ \bibnamefont
  {Huang}}, \bibinfo {author} {\bibfnamefont {J.}~\bibnamefont {Zhou}},\ and\
  \bibinfo {author} {\bibfnamefont {S.-Q.}\ \bibnamefont {Shen}},\ }\bibfield
  {title} {\bibinfo {title} {Topological responses from chiral anomaly in
  multi-Weyl semimetals},\ }\href {https://doi.org/10.1103/PhysRevB.96.085201}
  {\bibfield  {journal} {\bibinfo  {journal} {Phys. Rev. B}\ }\textbf {\bibinfo
  {volume} {96}},\ \bibinfo {pages} {085201} (\bibinfo {year}
  {2017})}\BibitemShut {NoStop}%
\bibitem [{\citenamefont {Bharti}\ and\ \citenamefont {Dixit}(2023)}]{Bharti}%
  \BibitemOpen
  \bibfield  {author} {\bibinfo {author} {\bibfnamefont {A.}~\bibnamefont
  {Bharti}}\ and\ \bibinfo {author} {\bibfnamefont {G.}~\bibnamefont {Dixit}},\
  }\bibfield  {title} {\bibinfo {title} {Role of topological charges in the
  nonlinear optical response from Weyl semimetals},\ }\href
  {https://doi.org/10.1103/PhysRevB.107.224308} {\bibfield  {journal} {\bibinfo
   {journal} {Phys. Rev. B}\ }\textbf {\bibinfo {volume} {107}},\ \bibinfo
  {pages} {224308} (\bibinfo {year} {2023})}\BibitemShut {NoStop}%
\bibitem [{\citenamefont {Ahn}\ \emph {et~al.}(2017)\citenamefont {Ahn},
  \citenamefont {Mele},\ and\ \citenamefont {Min}}]{Ahn}%
  \BibitemOpen
  \bibfield  {author} {\bibinfo {author} {\bibfnamefont {S.}~\bibnamefont
  {Ahn}}, \bibinfo {author} {\bibfnamefont {E.~J.}\ \bibnamefont {Mele}},\ and\
  \bibinfo {author} {\bibfnamefont {H.}~\bibnamefont {Min}},\ }\bibfield
  {title} {\bibinfo {title} {Optical conductivity of multi-Weyl semimetals},\
  }\href {https://doi.org/10.1103/physrevb.95.161112} {\bibfield  {journal}
  {\bibinfo  {journal} {Phys. Rev. B}\ }\textbf {\bibinfo {volume} {95}},\
  \bibinfo {pages} {161112} (\bibinfo {year} {2017})}\BibitemShut {NoStop}%
\bibitem [{\citenamefont {Mukherjee}\ and\ \citenamefont
  {Carbotte}(2018)}]{Mukherjee}%
  \BibitemOpen
  \bibfield  {author} {\bibinfo {author} {\bibfnamefont {S.~P.}\ \bibnamefont
  {Mukherjee}}\ and\ \bibinfo {author} {\bibfnamefont {J.~P.}\ \bibnamefont
  {Carbotte}},\ }\bibfield  {title} {\bibinfo {title} {Doping and tilting on
  optics in noncentrosymmetric multi-Weyl semimetals},\ }\href
  {https://doi.org/10.1103/physrevb.97.045150} {\bibfield  {journal} {\bibinfo
  {journal} {Phys. Rev. B}\ }\textbf {\bibinfo {volume} {97}},\ \bibinfo
  {pages} {045150} (\bibinfo {year} {2018})}\BibitemShut {NoStop}%
\bibitem [{\citenamefont {Park}\ \emph {et~al.}(2017)\citenamefont {Park},
  \citenamefont {Woo}, \citenamefont {Mele},\ and\ \citenamefont {Min}}]{Park}%
  \BibitemOpen
  \bibfield  {author} {\bibinfo {author} {\bibfnamefont {S.}~\bibnamefont
  {Park}}, \bibinfo {author} {\bibfnamefont {S.}~\bibnamefont {Woo}}, \bibinfo
  {author} {\bibfnamefont {E.~J.}\ \bibnamefont {Mele}},\ and\ \bibinfo
  {author} {\bibfnamefont {H.}~\bibnamefont {Min}},\ }\bibfield  {title}
  {\bibinfo {title} {Semiclassical boltzmann transport theory for multi-Weyl
  semimetals},\ }\href {https://doi.org/10.1103/physrevb.95.161113} {\bibfield
  {journal} {\bibinfo  {journal} {Phys. Rev. B}\ }\textbf {\bibinfo {volume}
  {95}},\ \bibinfo {pages} {161113} (\bibinfo {year} {2017})}\BibitemShut
  {NoStop}%
\bibitem [{\citenamefont {Silva}\ \emph {et~al.}(2022)\citenamefont {Silva},
  \citenamefont {Mizobata}, \citenamefont {Sanches}, \citenamefont {Ricco},
  \citenamefont {Shelykh}, \citenamefont {de~Souza}, \citenamefont {Figueira},
  \citenamefont {Vernek},\ and\ \citenamefont {Seridonio}}]{TopoFano}%
  \BibitemOpen
  \bibfield  {author} {\bibinfo {author} {\bibfnamefont {W.~C.}\ \bibnamefont
  {Silva}}, \bibinfo {author} {\bibfnamefont {W.~N.}\ \bibnamefont {Mizobata}},
  \bibinfo {author} {\bibfnamefont {J.~E.}\ \bibnamefont {Sanches}}, \bibinfo
  {author} {\bibfnamefont {L.~S.}\ \bibnamefont {Ricco}}, \bibinfo {author}
  {\bibfnamefont {I.~A.}\ \bibnamefont {Shelykh}}, \bibinfo {author}
  {\bibfnamefont {M.}~\bibnamefont {de~Souza}}, \bibinfo {author}
  {\bibfnamefont {M.~S.}\ \bibnamefont {Figueira}}, \bibinfo {author}
  {\bibfnamefont {E.}~\bibnamefont {Vernek}},\ and\ \bibinfo {author}
  {\bibfnamefont {A.~C.}\ \bibnamefont {Seridonio}},\ }\bibfield  {title}
  {\bibinfo {title} {Topological charge fano effect in multi-Weyl semimetals},\
  }\href {https://doi.org/10.1103/PhysRevB.105.235135} {\bibfield  {journal}
  {\bibinfo  {journal} {Phys. Rev. B}\ }\textbf {\bibinfo {volume} {105}},\
  \bibinfo {pages} {235135} (\bibinfo {year} {2022})}\BibitemShut {NoStop}%
\bibitem [{\citenamefont {Fu}\ and\ \citenamefont {Wang}(2022)}]{Thermo1}%
  \BibitemOpen
  \bibfield  {author} {\bibinfo {author} {\bibfnamefont {L.~X.}\ \bibnamefont
  {Fu}}\ and\ \bibinfo {author} {\bibfnamefont {C.~M.}\ \bibnamefont {Wang}},\
  }\bibfield  {title} {\bibinfo {title} {Thermoelectric transport of multi-Weyl
  semimetals in the quantum limit},\ }\href
  {https://doi.org/10.1103/PhysRevB.105.035201} {\bibfield  {journal} {\bibinfo
   {journal} {Phys. Rev. B}\ }\textbf {\bibinfo {volume} {105}},\ \bibinfo
  {pages} {035201} (\bibinfo {year} {2022})}\BibitemShut {NoStop}%
\bibitem [{\citenamefont {Gorbar}\ \emph {et~al.}(2017)\citenamefont {Gorbar},
  \citenamefont {Miransky}, \citenamefont {Shovkovy},\ and\ \citenamefont
  {Sukhachov}}]{Thermo2}%
  \BibitemOpen
  \bibfield  {author} {\bibinfo {author} {\bibfnamefont {E.~V.}\ \bibnamefont
  {Gorbar}}, \bibinfo {author} {\bibfnamefont {V.~A.}\ \bibnamefont
  {Miransky}}, \bibinfo {author} {\bibfnamefont {I.~A.}\ \bibnamefont
  {Shovkovy}},\ and\ \bibinfo {author} {\bibfnamefont {P.~O.}\ \bibnamefont
  {Sukhachov}},\ }\bibfield  {title} {\bibinfo {title} {Anomalous
  thermoelectric phenomena in lattice models of multi-Weyl semimetals},\ }\href
  {https://doi.org/10.1103/PhysRevB.96.155138} {\bibfield  {journal} {\bibinfo
  {journal} {Phys. Rev. B}\ }\textbf {\bibinfo {volume} {96}},\ \bibinfo
  {pages} {155138} (\bibinfo {year} {2017})}\BibitemShut {NoStop}%
\bibitem [{\citenamefont {Chen}\ and\ \citenamefont {Fiete}(2016)}]{Chen}%
  \BibitemOpen
  \bibfield  {author} {\bibinfo {author} {\bibfnamefont {Q.}~\bibnamefont
  {Chen}}\ and\ \bibinfo {author} {\bibfnamefont {G.~A.}\ \bibnamefont
  {Fiete}},\ }\bibfield  {title} {\bibinfo {title} {Thermoelectric transport in
  double-Weyl semimetals},\ }\href {https://doi.org/10.1103/physrevb.93.155125}
  {\bibfield  {journal} {\bibinfo  {journal} {Phys. Rev. B}\ }\textbf {\bibinfo
  {volume} {93}},\ \bibinfo {pages} {155125} (\bibinfo {year}
  {2016})}\BibitemShut {NoStop}%
\bibitem [{\citenamefont {Xu}\ \emph {et~al.}(2011)\citenamefont {Xu},
  \citenamefont {Weng}, \citenamefont {Wang}, \citenamefont {Dai},\ and\
  \citenamefont {Fang}}]{Xu}%
  \BibitemOpen
  \bibfield  {author} {\bibinfo {author} {\bibfnamefont {G.}~\bibnamefont
  {Xu}}, \bibinfo {author} {\bibfnamefont {H.}~\bibnamefont {Weng}}, \bibinfo
  {author} {\bibfnamefont {Z.}~\bibnamefont {Wang}}, \bibinfo {author}
  {\bibfnamefont {X.}~\bibnamefont {Dai}},\ and\ \bibinfo {author}
  {\bibfnamefont {Z.}~\bibnamefont {Fang}},\ }\bibfield  {title} {\bibinfo
  {title} {Chern semimetal and the quantized anomalous Hall effect in
  HgCr2Se4},\ }\href {https://doi.org/10.1103/physrevlett.107.186806}
  {\bibfield  {journal} {\bibinfo  {journal} {Phys. Rev. Lett.}\ }\textbf
  {\bibinfo {volume} {107}},\ \bibinfo {pages} {186806} (\bibinfo {year}
  {2011})}\BibitemShut {NoStop}%
\bibitem [{\citenamefont {Xiong}\ \emph {et~al.}(2022)\citenamefont {Xiong},
  \citenamefont {Honerkamp}, \citenamefont {Kennes},\ and\ \citenamefont
  {Nag}}]{Xiong}%
  \BibitemOpen
  \bibfield  {author} {\bibinfo {author} {\bibfnamefont {F.}~\bibnamefont
  {Xiong}}, \bibinfo {author} {\bibfnamefont {C.}~\bibnamefont {Honerkamp}},
  \bibinfo {author} {\bibfnamefont {D.~M.}\ \bibnamefont {Kennes}},\ and\
  \bibinfo {author} {\bibfnamefont {T.}~\bibnamefont {Nag}},\ }\bibfield
  {title} {\bibinfo {title} {Understanding the three-dimensional quantum Hall
  effect in generic multi-Weyl semimetals},\ }\href
  {https://doi.org/10.1103/PhysRevB.106.045424} {\bibfield  {journal} {\bibinfo
   {journal} {Phys. Rev. B}\ }\textbf {\bibinfo {volume} {106}},\ \bibinfo
  {pages} {045424} (\bibinfo {year} {2022})}\BibitemShut {NoStop}%
\bibitem [{\citenamefont {Dantas}\ \emph {et~al.}(2018)\citenamefont {Dantas},
  \citenamefont {Pe\~na Benitez}, \citenamefont {Roy},\ and\ \citenamefont
  {Sur\'owka}}]{Dantas2018}%
  \BibitemOpen
  \bibfield  {author} {\bibinfo {author} {\bibfnamefont {R.~M.~A.}\
  \bibnamefont {Dantas}}, \bibinfo {author} {\bibfnamefont {F.}~\bibnamefont
  {Pe\~na Benitez}}, \bibinfo {author} {\bibfnamefont {B.}~\bibnamefont
  {Roy}},\ and\ \bibinfo {author} {\bibfnamefont {P.}~\bibnamefont
  {Sur\'owka}},\ }\bibfield  {title} {\bibinfo {title} {Magnetotransport in
  multi-Weyl semimetals: a kinetic theory approach},\ }\href
  {https://doi.org/10.1007/jhep12(2018)069} {\bibfield  {journal} {\bibinfo
  {journal} {Journal of High Energy Physics}\ }\textbf {\bibinfo {volume}
  {2018}},\ \bibinfo {pages} {69} (\bibinfo {year} {2018})}\BibitemShut
  {NoStop}%
\bibitem [{\citenamefont {Ghosh}\ \emph
  {et~al.}(2024{\natexlab{a}})\citenamefont {Ghosh}, \citenamefont {Nandy},
  \citenamefont {Zhu},\ and\ \citenamefont {Taraphder}}]{Transport}%
  \BibitemOpen
  \bibfield  {author} {\bibinfo {author} {\bibfnamefont {S.}~\bibnamefont
  {Ghosh}}, \bibinfo {author} {\bibfnamefont {S.}~\bibnamefont {Nandy}},
  \bibinfo {author} {\bibfnamefont {J.-X.}\ \bibnamefont {Zhu}},\ and\ \bibinfo
  {author} {\bibfnamefont {A.}~\bibnamefont {Taraphder}},\ }\bibfield  {title}
  {\bibinfo {title} {Signature of nodal topology in nonlinear quantum transport
  across junctions in Weyl and multi-Weyl semimetals},\ }\href
  {https://doi.org/10.1103/PhysRevB.109.045437} {\bibfield  {journal} {\bibinfo
   {journal} {Phys. Rev. B}\ }\textbf {\bibinfo {volume} {109}},\ \bibinfo
  {pages} {045437} (\bibinfo {year} {2024}{\natexlab{a}})}\BibitemShut
  {NoStop}%
\bibitem [{\citenamefont {Ghosh}\ \emph
  {et~al.}(2024{\natexlab{b}})\citenamefont {Ghosh}, \citenamefont {Nandy},
  \citenamefont {Zhu},\ and\ \citenamefont {Taraphder}}]{NonlinearQT}%
  \BibitemOpen
  \bibfield  {author} {\bibinfo {author} {\bibfnamefont {S.}~\bibnamefont
  {Ghosh}}, \bibinfo {author} {\bibfnamefont {S.}~\bibnamefont {Nandy}},
  \bibinfo {author} {\bibfnamefont {J.-X.}\ \bibnamefont {Zhu}},\ and\ \bibinfo
  {author} {\bibfnamefont {A.}~\bibnamefont {Taraphder}},\ }\bibfield  {title}
  {\bibinfo {title} {Signature of nodal topology in nonlinear quantum transport
  across junctions in Weyl and multi-Weyl semimetals},\ }\href
  {https://doi.org/10.1103/PhysRevB.109.045437} {\bibfield  {journal} {\bibinfo
   {journal} {Phys. Rev. B}\ }\textbf {\bibinfo {volume} {109}},\ \bibinfo
  {pages} {045437} (\bibinfo {year} {2024}{\natexlab{b}})}\BibitemShut
  {NoStop}%
\bibitem [{\citenamefont {Pedrosa}\ \emph {et~al.}(2021)\citenamefont
  {Pedrosa}, \citenamefont {Silva},\ and\ \citenamefont {Vernek}}]{Pedrosa}%
  \BibitemOpen
  \bibfield  {author} {\bibinfo {author} {\bibfnamefont {G.~T.~D.}\
  \bibnamefont {Pedrosa}}, \bibinfo {author} {\bibfnamefont {J.~F.}\
  \bibnamefont {Silva}},\ and\ \bibinfo {author} {\bibfnamefont
  {E.}~\bibnamefont {Vernek}},\ }\bibfield  {title} {\bibinfo {title} {Kondo
  screening regimes in multi-Dirac and Weyl systems},\ }\href
  {https://doi.org/10.1103/physrevb.103.045137} {\bibfield  {journal} {\bibinfo
   {journal} {Phys. Rev. B}\ }\textbf {\bibinfo {volume} {103}},\ \bibinfo
  {pages} {045137} (\bibinfo {year} {2021})}\BibitemShut {NoStop}%
\bibitem [{\citenamefont {Silva}\ and\ \citenamefont
  {Miranda}(2024)}]{Miranda}%
  \BibitemOpen
  \bibfield  {author} {\bibinfo {author} {\bibfnamefont {J.~F.}\ \bibnamefont
  {Silva}}\ and\ \bibinfo {author} {\bibfnamefont {E.}~\bibnamefont
  {Miranda}},\ }\bibfield  {title} {\bibinfo {title} {Multi-Dirac and Weyl
  physics in heavy-fermion systems},\ }\href
  {https://doi.org/10.1103/PhysRevB.109.035153} {\bibfield  {journal} {\bibinfo
   {journal} {Phys. Rev. B}\ }\textbf {\bibinfo {volume} {109}},\ \bibinfo
  {pages} {035153} (\bibinfo {year} {2024})}\BibitemShut {NoStop}%
\bibitem [{\citenamefont {Kofuji}\ \emph {et~al.}(2021)\citenamefont {Kofuji},
  \citenamefont {Michishita},\ and\ \citenamefont {Peters}}]{KondoLattice1}%
  \BibitemOpen
  \bibfield  {author} {\bibinfo {author} {\bibfnamefont {A.}~\bibnamefont
  {Kofuji}}, \bibinfo {author} {\bibfnamefont {Y.}~\bibnamefont {Michishita}},\
  and\ \bibinfo {author} {\bibfnamefont {R.}~\bibnamefont {Peters}},\
  }\bibfield  {title} {\bibinfo {title} {Effects of strong correlations on the
  nonlinear response in Weyl-Kondo semimetals},\ }\href
  {https://doi.org/10.1103/PhysRevB.104.085151} {\bibfield  {journal} {\bibinfo
   {journal} {Phys. Rev. B}\ }\textbf {\bibinfo {volume} {104}},\ \bibinfo
  {pages} {085151} (\bibinfo {year} {2021})}\BibitemShut {NoStop}%
\bibitem [{\citenamefont {Lu}\ \emph {et~al.}(2019)\citenamefont {Lu},
  \citenamefont {Chou}, \citenamefont {Chung},\ and\ \citenamefont
  {Mou}}]{KondoLattice3}%
  \BibitemOpen
  \bibfield  {author} {\bibinfo {author} {\bibfnamefont {Y.-W.}\ \bibnamefont
  {Lu}}, \bibinfo {author} {\bibfnamefont {P.-H.}\ \bibnamefont {Chou}},
  \bibinfo {author} {\bibfnamefont {C.-H.}\ \bibnamefont {Chung}},\ and\
  \bibinfo {author} {\bibfnamefont {C.-Y.}\ \bibnamefont {Mou}},\ }\bibfield
  {title} {\bibinfo {title} {Tunable topological semimetallic phases in Kondo
  lattice systems},\ }\href {https://doi.org/10.1103/PhysRevB.99.035141}
  {\bibfield  {journal} {\bibinfo  {journal} {Phys. Rev. B}\ }\textbf {\bibinfo
  {volume} {99}},\ \bibinfo {pages} {035141} (\bibinfo {year}
  {2019})}\BibitemShut {NoStop}%
\bibitem [{\citenamefont {Sun}\ and\ \citenamefont {Wang}(2017)}]{RKKY1}%
  \BibitemOpen
  \bibfield  {author} {\bibinfo {author} {\bibfnamefont {Y.}~\bibnamefont
  {Sun}}\ and\ \bibinfo {author} {\bibfnamefont {A.}~\bibnamefont {Wang}},\
  }\bibfield  {title} {\bibinfo {title} {RKKY interaction of magnetic
  impurities in multi-Weyl semimetals},\ }\href
  {https://doi.org/10.1088/1361-648x/aa8932} {\bibfield  {journal} {\bibinfo
  {journal} {Journal of Physics: Condensed Matter}\ }\textbf {\bibinfo {volume}
  {29}},\ \bibinfo {pages} {435306} (\bibinfo {year} {2017})}\BibitemShut
  {NoStop}%
\bibitem [{\citenamefont {Chang}\ \emph {et~al.}(2015)\citenamefont {Chang},
  \citenamefont {Zhou}, \citenamefont {Wang}, \citenamefont {Shan},\ and\
  \citenamefont {Xiao}}]{RKKY2}%
  \BibitemOpen
  \bibfield  {author} {\bibinfo {author} {\bibfnamefont {H.-R.}\ \bibnamefont
  {Chang}}, \bibinfo {author} {\bibfnamefont {J.}~\bibnamefont {Zhou}},
  \bibinfo {author} {\bibfnamefont {S.-X.}\ \bibnamefont {Wang}}, \bibinfo
  {author} {\bibfnamefont {W.-Y.}\ \bibnamefont {Shan}},\ and\ \bibinfo
  {author} {\bibfnamefont {D.}~\bibnamefont {Xiao}},\ }\bibfield  {title}
  {\bibinfo {title} {RKKY interaction of magnetic impurities in Dirac and Weyl
  semimetals},\ }\href {https://doi.org/10.1103/PhysRevB.92.241103} {\bibfield
  {journal} {\bibinfo  {journal} {Phys. Rev. B}\ }\textbf {\bibinfo {volume}
  {92}},\ \bibinfo {pages} {241103} (\bibinfo {year} {2015})}\BibitemShut
  {NoStop}%
\bibitem [{\citenamefont {Verma}\ \emph {et~al.}(2020)\citenamefont {Verma},
  \citenamefont {Giri}, \citenamefont {Fertig},\ and\ \citenamefont
  {Kundu}}]{RKKY3}%
  \BibitemOpen
  \bibfield  {author} {\bibinfo {author} {\bibfnamefont {S.}~\bibnamefont
  {Verma}}, \bibinfo {author} {\bibfnamefont {D.}~\bibnamefont {Giri}},
  \bibinfo {author} {\bibfnamefont {H.~A.}\ \bibnamefont {Fertig}},\ and\
  \bibinfo {author} {\bibfnamefont {A.}~\bibnamefont {Kundu}},\ }\bibfield
  {title} {\bibinfo {title} {RKKY coupling in Weyl semimetal thin films},\
  }\href {https://doi.org/10.1103/PhysRevB.101.085419} {\bibfield  {journal}
  {\bibinfo  {journal} {Phys. Rev. B}\ }\textbf {\bibinfo {volume} {101}},\
  \bibinfo {pages} {085419} (\bibinfo {year} {2020})}\BibitemShut {NoStop}%
\bibitem [{\citenamefont {Paul}\ \emph {et~al.}(2021)\citenamefont {Paul},
  \citenamefont {Islam}, \citenamefont {Dutta},\ and\ \citenamefont
  {Saha}}]{RKKY4}%
  \BibitemOpen
  \bibfield  {author} {\bibinfo {author} {\bibfnamefont {G.~C.}\ \bibnamefont
  {Paul}}, \bibinfo {author} {\bibfnamefont {S.~F.}\ \bibnamefont {Islam}},
  \bibinfo {author} {\bibfnamefont {P.}~\bibnamefont {Dutta}},\ and\ \bibinfo
  {author} {\bibfnamefont {A.}~\bibnamefont {Saha}},\ }\bibfield  {title}
  {\bibinfo {title} {Signatures of interfacial topological chiral modes via
  RKKY exchange interaction in Dirac and Weyl systems},\ }\href
  {https://doi.org/10.1103/PhysRevB.103.115306} {\bibfield  {journal} {\bibinfo
   {journal} {Phys. Rev. B}\ }\textbf {\bibinfo {volume} {103}},\ \bibinfo
  {pages} {115306} (\bibinfo {year} {2021})}\BibitemShut {NoStop}%
\bibitem [{\citenamefont {Duan}\ \emph {et~al.}(2024)\citenamefont {Duan},
  \citenamefont {Cai}, \citenamefont {Wei}, \citenamefont {Chen}, \citenamefont
  {Wu}, \citenamefont {Deng}, \citenamefont {Wang},\ and\ \citenamefont
  {Yang}}]{RKKY5}%
  \BibitemOpen
  \bibfield  {author} {\bibinfo {author} {\bibfnamefont {H.-J.}\ \bibnamefont
  {Duan}}, \bibinfo {author} {\bibfnamefont {S.-M.}\ \bibnamefont {Cai}},
  \bibinfo {author} {\bibfnamefont {X.}~\bibnamefont {Wei}}, \bibinfo {author}
  {\bibfnamefont {Y.-C.}\ \bibnamefont {Chen}}, \bibinfo {author}
  {\bibfnamefont {Y.-J.}\ \bibnamefont {Wu}}, \bibinfo {author} {\bibfnamefont
  {M.-X.}\ \bibnamefont {Deng}}, \bibinfo {author} {\bibfnamefont
  {R.}~\bibnamefont {Wang}},\ and\ \bibinfo {author} {\bibfnamefont
  {M.}~\bibnamefont {Yang}},\ }\bibfield  {title} {\bibinfo {title} {RKKY
  signals characterizing the topological phase transitions in floquet Dirac
  semimetals},\ }\href {https://doi.org/10.1103/PhysRevB.109.205149} {\bibfield
   {journal} {\bibinfo  {journal} {Phys. Rev. B}\ }\textbf {\bibinfo {volume}
  {109}},\ \bibinfo {pages} {205149} (\bibinfo {year} {2024})}\BibitemShut
  {NoStop}%
\bibitem [{\citenamefont {Hubbard}(1963)}]{Hubbard1963}%
  \BibitemOpen
  \bibfield  {author} {\bibinfo {author} {\bibfnamefont {J.}~\bibnamefont
  {Hubbard}},\ }\bibfield  {title} {\bibinfo {title} {Electron correlations in
  narrow energy bands},\ }\href {https://doi.org/10.1098/rspa.1963.0200}
  {\bibfield  {journal} {\bibinfo  {journal} {Proceedings of the Royal Society
  of London. Series A, Mathematical and Physical Sciences}\ }\textbf {\bibinfo
  {volume} {276}},\ \bibinfo {pages} {238} (\bibinfo {year}
  {1963})}\BibitemShut {NoStop}%
\bibitem [{\citenamefont {Bruus}\ and\ \citenamefont
  {Flensberg}(2012)}]{Flensberg}%
  \BibitemOpen
  \bibfield  {author} {\bibinfo {author} {\bibfnamefont {H.}~\bibnamefont
  {Bruus}}\ and\ \bibinfo {author} {\bibfnamefont {K.}~\bibnamefont
  {Flensberg}},\ }\bibfield  {title} {\bibinfo {title} {Many-body quantum
  theory in condensed matter physics, an introduction},\ }\href@noop {}
  {\bibfield  {journal} {\bibinfo  {journal} {(Oxford: Oxford University
  Press)}\ } (\bibinfo {year} {2012})}\BibitemShut {NoStop}%
\bibitem [{\citenamefont {Guessi}\ \emph {et~al.}(2015)\citenamefont {Guessi},
  \citenamefont {Machado}, \citenamefont {Marques}, \citenamefont {Ricco},
  \citenamefont {Kristinsson}, \citenamefont {Yoshida}, \citenamefont
  {Shelykh}, \citenamefont {de~Souza},\ and\ \citenamefont
  {Seridonio}}]{HubbardI}%
  \BibitemOpen
  \bibfield  {author} {\bibinfo {author} {\bibfnamefont {L.~H.}\ \bibnamefont
  {Guessi}}, \bibinfo {author} {\bibfnamefont {R.~S.}\ \bibnamefont {Machado}},
  \bibinfo {author} {\bibfnamefont {Y.}~\bibnamefont {Marques}}, \bibinfo
  {author} {\bibfnamefont {L.~S.}\ \bibnamefont {Ricco}}, \bibinfo {author}
  {\bibfnamefont {K.}~\bibnamefont {Kristinsson}}, \bibinfo {author}
  {\bibfnamefont {M.}~\bibnamefont {Yoshida}}, \bibinfo {author} {\bibfnamefont
  {I.~A.}\ \bibnamefont {Shelykh}}, \bibinfo {author} {\bibfnamefont
  {M.}~\bibnamefont {de~Souza}},\ and\ \bibinfo {author} {\bibfnamefont
  {A.~C.}\ \bibnamefont {Seridonio}},\ }\bibfield  {title} {\bibinfo {title}
  {Catching the bound states in the continuum of a phantom atom in graphene},\
  }\href {https://doi.org/10.1103/PhysRevB.92.045409} {\bibfield  {journal}
  {\bibinfo  {journal} {Phys. Rev. B}\ }\textbf {\bibinfo {volume} {92}},\
  \bibinfo {pages} {045409} (\bibinfo {year} {2015})}\BibitemShut {NoStop}%
\bibitem [{\citenamefont {Anderson}(1961)}]{Anderson}%
  \BibitemOpen
  \bibfield  {author} {\bibinfo {author} {\bibfnamefont {P.~W.}\ \bibnamefont
  {Anderson}},\ }\bibfield  {title} {\bibinfo {title} {Localized magnetic
  states in metals},\ }\href {https://doi.org/10.1103/PhysRev.124.41}
  {\bibfield  {journal} {\bibinfo  {journal} {Phys. Rev.}\ }\textbf {\bibinfo
  {volume} {124}},\ \bibinfo {pages} {41} (\bibinfo {year} {1961})}\BibitemShut
  {NoStop}%
\bibitem [{\citenamefont {Mitchell}\ and\ \citenamefont
  {Fritz}(2015)}]{KondoWeyl1}%
  \BibitemOpen
  \bibfield  {author} {\bibinfo {author} {\bibfnamefont {A.~K.}\ \bibnamefont
  {Mitchell}}\ and\ \bibinfo {author} {\bibfnamefont {L.}~\bibnamefont
  {Fritz}},\ }\bibfield  {title} {\bibinfo {title} {Kondo effect in
  three-dimensional Dirac and Weyl systems},\ }\href
  {https://doi.org/10.1103/PhysRevB.92.121109} {\bibfield  {journal} {\bibinfo
  {journal} {Phys. Rev. B}\ }\textbf {\bibinfo {volume} {92}},\ \bibinfo
  {pages} {121109} (\bibinfo {year} {2015})}\BibitemShut {NoStop}%
\end{thebibliography}
%
\end{document}